\documentclass[conference]{IEEEtran}
\IEEEoverridecommandlockouts
\usepackage{cite}
\usepackage{amsmath,amssymb,amsfonts}
\usepackage{algorithmic}
\usepackage[ruled,vlined]{algorithm2e}
\usepackage{graphicx}
\usepackage{textcomp}
\usepackage{xcolor}
\usepackage{url}
\usepackage{footnote}
\usepackage[perpage]{footmisc}
\usepackage{balance}
\usepackage{multirow}

\def\BibTeX{{\rm B\kern-.05em{\sc i\kern-.025em b}\kern-.08em
    T\kern-.1667em\lower.7ex\hbox{E}\kern-.125emX}}
\begin{document}

\newcommand{\eg}{{\textit{e.g., }}}
\newcommand{\ie}{{\textit{i.e., }}}
\newcommand{\al}{\textit{et al. }}
\newcommand{\Foutse}[1]{\textcolor{red}{{\it [Foutse: #1]}}}
\newcommand{\Amin}[1]{\textcolor{blue}{{\it [Amin: #1]}}}

\title{Design Smells in Deep Learning Programs: An Empirical Study\\
}


\author{\IEEEauthorblockN{Amin Nikanjam, Foutse Khomh}
\IEEEauthorblockA{\textit{SWAT Lab., Polytechnique Montréal, Montréal, Canada} \\
\{amin.nikanjam, foutse.khomh\}@polymtl.ca}
}

\maketitle

\begin{abstract}
Nowadays, we are witnessing an increasing adoption of Deep Learning (DL) based software systems in many industries. Designing a DL program requires constructing a deep neural network (DNN) and then training it on a dataset. 
This process requires that developers make multiple architectural (e.g., type, size, number, and order of layers) and configuration (e.g., optimizer, regularization methods, and activation functions) choices that affect the quality of the DL models, and consequently software quality.  
An under-specified or poorly-designed DL model may train successfully but is likely to perform poorly when deployed in production. Design smells in DL programs are poor design and--or configuration decisions taken during the development of DL components, that are likely to have a negative impact on the performance (i.e., prediction accuracy) and then quality of DL based software systems. 
In this paper, we present a catalogue of 8 design smells for a popular DL architecture, namely 
deep Feedforward Neural Networks which is widely employed in industrial applications. The design smells were identified through a review of the existing literature on DL design and a manual inspection of 659 DL programs with performance issues and design inefficiencies. 
The smells are specified by describing their context, consequences, and recommended refactorings. To provide empirical evidence on the relevance and perceived impact of the proposed design smells, we conducted a survey with 81 DL developers. In general, the developers perceived the proposed design smells as reflective of design or implementation problems, with agreement levels varying between 47\% and 68\%.

\end{abstract}


\section{Introduction}
\begin{figure*}
    \centering
    \includegraphics[width=.9\textwidth]{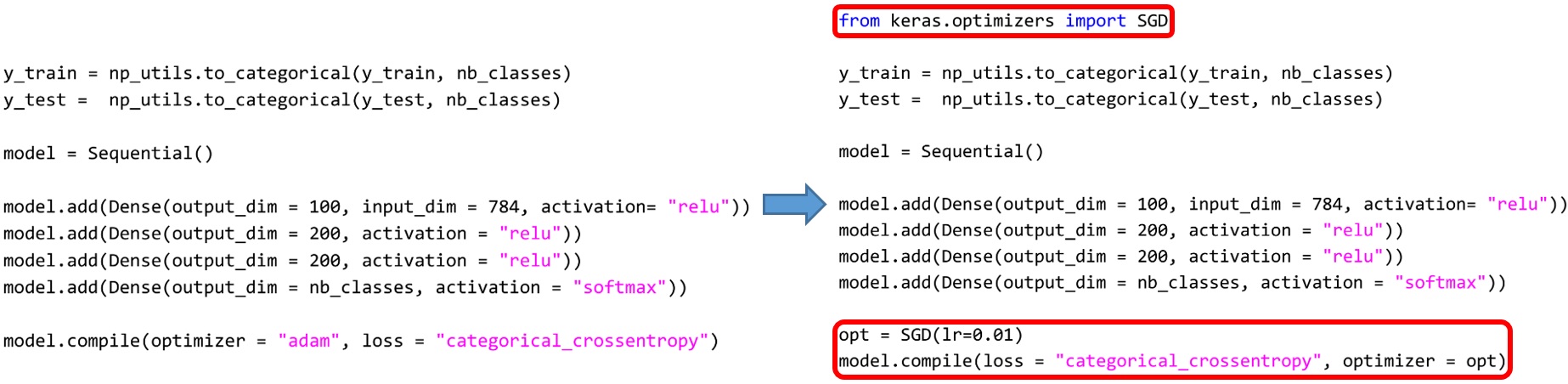}
    \caption{A poorly-designed model (left) and its refactored version (right). The optimizer has been changed to improve the performance in a classification problem. The recommended changes have been highlighted by the red color (simplified from SO\_37213388).}
    \label{fig:motivation}
\end{figure*}

Nowadays, we are observing an increasing deployment of software systems based on Deep Learning (DL) in real life, from personal banking to autonomous driving \cite{heaton2020applications}. A DL program encodes the network structure of a desirable DL model and the process by which the model learns from a training dataset. Easy-to-use libraries such as Keras have been introduced to simplify the development process of DL programs. However, leveraging these libraries to implement a DL program is still challenging, in particular for developers who are not experts in Machine Learning (ML) and neural networks. A developer must make multiple architectural (e.g., type, size, number, and order of layers) and configuration (e.g., optimizer, regularization methods, and activation functions) choices that affect the quality of the DL models, and consequently software quality. 
A poorly-designed DL model may train successfully but is likely to perform poorly when deployed in production. 
Design smells in DL programs are poor design and–or configuration decisions that can have a negative impact on the performance and then quality of a DL-based software system. By performance, we mean accuracy of prediction, like precision of classifying samples in the correct target class, that may affect the quality of final decisions. In software engineering, traditionally code/design smells deal with non-functional requirements such as testability or maintainability, but in ML-based systems the accuracy can be regarded as a functional requirement. In this paper, we define design smells in DL programs as poorly designed/configured models that may affect the entire performance, i.e. prediction accuracy, of DL-based systems. An example of a poor design decision in a DL model and its refactored version are shown in Fig. \ref{fig:motivation}. 
When training the 
model to detect images of handwritten digits, the developer selected an inadequate optimiser at the last line; i.e., ``Adam" in \texttt{compile} function instead of Stochastic Gradient Descent (SGD) optimizer as pointed in the correct answer, which caused the accuracy of the model to remained unchanged between epochs 2 to 10. 
Consequently, the model was not able to train well on the data, leading to a low classification accuracy. Such low classification accuracy results in poor decisions like misclassification of input images. 
Changing the optimizer led to successfully addressing the problem and the performance improved significantly. 

Deploying a DL model with poor performance can have severe consequences, especially in the context of safety-critical systems. It is therefore important to raise the awareness of development teams about poor design and configuration issues that are likely to have a negative impact on the quality of DL models. 
Design smells can cause a 
program to exhibit extraordinary poor accuracy or other low quality outputs during the execution phase. 
Having a list of known bad design practices for DL models can help developers avoid pitfalls during the development of their DL programs; resulting in better software quality. Although poor design choices and performance issues in DL programs have been studied previously \cite{DL_bugs_1, DL_bugs_2, CNN_design_patterns, CNN_principles}, to the best of our knowledge, this paper is the first empirical study on design smells in DL programs. 

\begin{figure*}
    \centering
    \includegraphics[width=.6\textwidth]{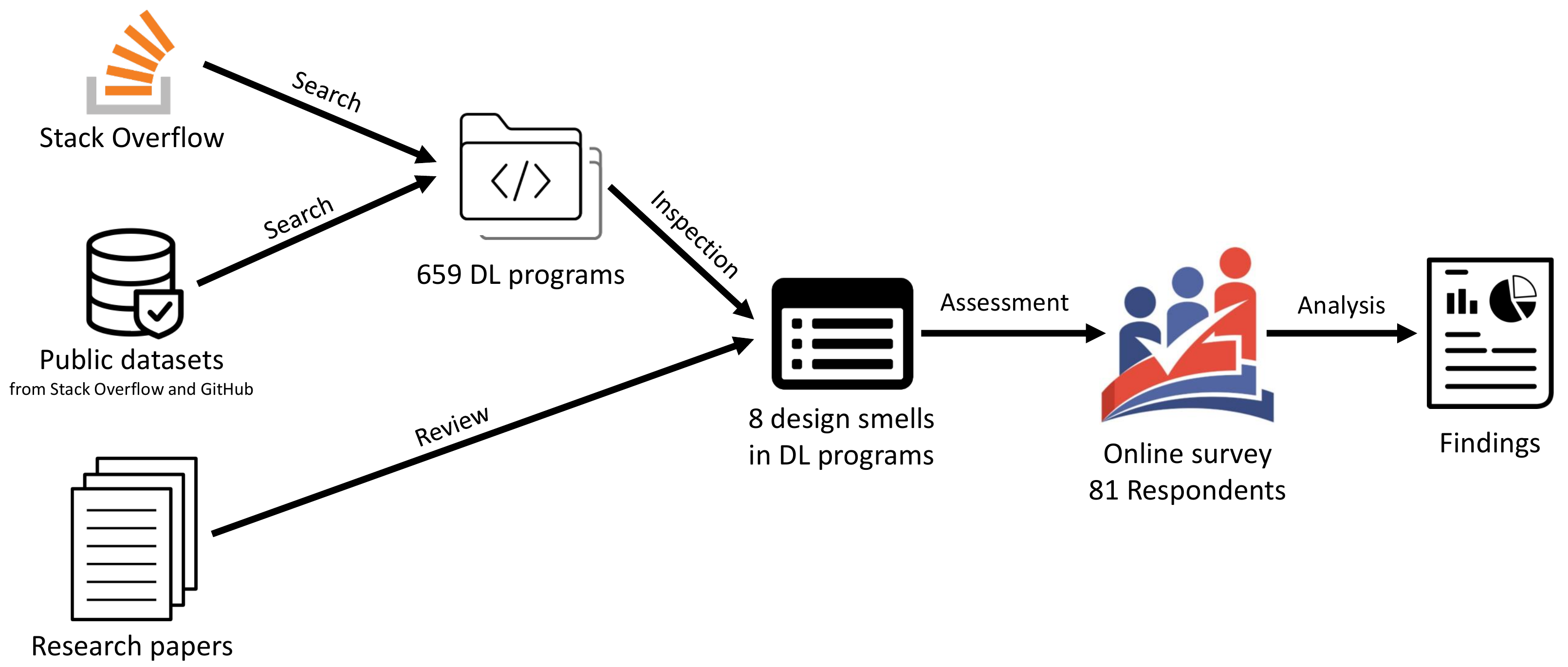}
    \caption{Schematic diagram of our study.}
    \label{fig:metodology}
\end{figure*}

In this paper, we propose a catalog of 8 design smells in DL models with a focus on deep Feedforward Neural Networks (FNN) that use convolutional components. Fig. \ref{fig:metodology} illustrates the schematic diagram of our study in this paper. We start by conducting an investigation to determine the type of smells and their prevalence using two main sources: (1) previous research studies that highlighted bad practices in designing DL models, and (2) DL programs with design or performance issues. We have identified two main categories of design smells: Formation of the feature map and usage of regularization methods. Context, consequences and recommended refactorings for removing each smell are specified in the catalogue with some examples from real DL programs. Finally, the relevance of design smells are assessed by running a survey among 81 eligible DL developers/researchers. In general, the developers perceived the proposed design smells as reflective of design or implementation problems, with agreement levels varying between 47\% and 68\%.
The contributions of this paper are: 1) proposing a catalogue of 8 design smells in DL models, and 2) validating the catalogue through a survey with 81 eligible DL developers/researchers.

The remainder of this paper is organised as follows. Section~\ref{background} briefly reviews background knowledge about DL, deep FNNs and the development of DL program/models. Section \ref{smells} introduces the methodology adopted for the identification of smells and a full description of the identified design smells in DL models. Section~\ref{survey} presents the design of the survey used 
to validate the proposed design smells, and the obtained results. Section~\ref{threats} discusses threats to the validity of this study. Finally, we conclude the paper and discuss future work in Section~\ref{conclusion}.
\section{Background}\label{background}
\subsection{Feedforward Neural Networks (FNN)}
FNN \cite{DL_ebook_2016} is the principal neural network architecture used for solving classification and function approximation problems, where the task is to learn a mapping function capable of converting input data to a target output. FNN consists of several, and sometimes diverse, sequences of layers of computational units. These computational layers are trained to extract features hierarchically. This starts from low-level features in early layers to high-level ones in middle layers. FNN, then, detects discriminative and informative patterns in last layers, which serve it to derive either the class label (in classification problems) or continuous outcome (in function approximation problems). It is called feedforward because the information flows in a forward manner from the input layer, through the hidden layers and to the output layer, e.g., a class probability or a predicted real value. The basic FNN architecture consists of stacking dense layers, where all the neurons of two consecutive layers are fully-connected.

The regularization is required to improve the convergence and generalizability of the training procedure of DNNs. Many regularization techniques have been proposed and the most used ones are dropout and batch normalisation (batchnorm). Dropout \cite{dropout} masks at every training iteration a random subset of units (i.e., nullify them). The stochasticity injected into the inference calculation, only during the training, prevents the co-adaptation of feature detectors and encourages the DNN to learn robust patterns against partially-hidden information. Batchnorm \cite{batchnorm} acts differently on activations by normalizing their values using statistics (i.e., mean and variance) of the current batch of data during the training. During the testing, it updates internally, the population statistics of all batches for each level of activations in order to switch to normalizing against population, rather than batch, statistics. This normalization of intermediary inputs data has shown its effectiveness in smoothing the loss landscape, which ensures faster and safer training convergence with high potential to escape weak local minima.

Convolutional architectures represent a particular type of FNN designed for multi-dimensional input data, such as 2D images, audio spectrograms, or 3D videos \cite{krizhevsky2012imagenet}. The benefit of Convolutional Neural Networks (CNN) lies in their ability to take into account the spatial information in their feature extraction process. To do that, CNNs stack, earlier, two specialized layers: 
\begin{itemize}
\item Convolutional layer: it applies spatial filters over the input data and each filter’s weights are learned to detect relevant features supporting the network’s task. Thus, it yields a feature map for each learned filter, where each unit is connected to a local region (i.e., size of spatial filtering window) in its previous layer’s feature maps.
\item Pooling layer: this layer performs spatial pooling over the computed feature map to reduce its dimensionality and retain the most relevant information. The spatial pooling can be either average or max aggregation that computes, respectively, the average or max of all the units in the specified spatial window.
\end{itemize}

Indeed, some bad configurations and poor design choices may definitely introduce inefficiencies on the internal functioning of the FNN or one of its components, which can hinder the expressiveness of mapping functions or computational resource consumption. Such configurations or design choices have been reported in several studies as a root cause of bad performance in DL programs \cite{DL_bugs_1, DL_bugs_2}. DL researchers have studied performance issues in DL models \cite{CNN_design_patterns, CNN_principles} as well. Moreover, other researchers have reported some principles and best practices for designing CNN \cite{systematic_CNNs, practical_CNNs}. 

\subsection{Developing DL programs}
The development of DL programs lies in constructing the Deep Neural Network (DNN) by calling built-in DL routines to create layers (processing units), then connecting them by either feeding one or more layers’ outputs as inputs to another. Then, the developer should train the DNN by configuring a learning algorithm on a dataset. The training process consists in updating iteratively the DNN’s parameters, towards minimizing the loss of DNN's predictions compared to the training data. A loss/cost function is defined to estimate the average distance between predicted and actual outcomes. Commonly, the best-fitted FNN is found after multiple epochs (i.e., passes over all the training data).

However, leveraging DL libraries to implement a DNN and then a training program for the designed DNN is not straightforward and it can be error-prone. DL libraries often have to trade off between the coverage of novel DL functionalities and the ease of rapid implementation and extension of DNN software prototypes. As a compromise solution, they uniformly include, for each newly-implemented DL functionality, a bundle of automated steps and default settings following its common usage trends. This enables quick prototyping of regular DNNs while keeping the flexibility to try other configurations with the tweakable setting options available for every provided DL routine. As a consequence, DL developers should be aware of the intricacies of these DL libraries to choose the appropriate configurations and avoid breaking their implicit assumptions in regard to the usage of their built-in routines.

\section{Design Smells in DL models}\label{smells}
In this section, first we describe our methodology for eliciting design smells by analyzing existing literature and related DL programs. Then, we explain identified design smells in feedforward DL models in detail. We explain the context of each smell, its characteristics, consequences, and the recommended refactoring to address it, following the template provided by Brown et al. \cite{brown1998refactoring}. Moreover, code snippets are provided as examples in some cases.

\subsection{Methodology}
In this study, we focus specifically on FNNs. This popular architecture inside the DL community is considered as “quintessential” in DL and they has many industrial applications like object recognition from images \cite{DL_ebook_2016}. In fact, a special feedforward architecture which is called Convolutional Neural Network (CNN) has shown its effectiveness on public computer vision datasets and competitions such as ImageNet classification \cite{deng2009imagenet} or COCO object detection \cite{lin2014microsoft}. Moreover, FNN is a conceptual milestone on the road to recurrent networks that are employed widely in Natural Language applications. Thus, we limit our study to deep FNNs and do not consider other DL models. 

The goal of this study is to identify 
design smells that could affect the performance 
of a DL program. 
We examined two main sources of information to identify such smells: (1) previous research studies that highlighted performance issues in DL models, and (2) 
DL programs that exhibited design or performance issues. We reviewed empirical research studies on DNN design principles and bad performance in DL programs to identify frequent and influential design smells in deep FNNs, including poor design choices/configurations that lead to bad performance in DL programs \cite{DL_bugs_1, DL_bugs_2}, performance issues in DL models \cite{CNN_design_patterns, CNN_principles}, and reported principles and best practices for designing CNN \cite{systematic_CNNs, practical_CNNs}. 

The second source of information about design smells is real DL programs that have design inefficiencies. To find a proper set of real-world design smells in DL programs, we have used two main sources: 1) samples found by directly searching over SO with keywords related to such issues, and 2) public datasets of faulty DL programs (from SO and GitHub) released by previous research studies. For the former, we chose SO because it is the most popular Q\&A forum for software development and has been leveraged by previous studies on DL software systems~\cite{DL_bugs_1, DL_bugs_2, DL_challenges}. Since TensorFlow and Keras are very popular among DL developers, in this paper we searched SO posts tagged by one of these libraries with the objective of collecting relevant DL models/programs. We refined our search queries with keywords related to the scope of our study: 
“low performance”, “bad performance” and “design issues”. 
We consider SO posts, containing full code scripts or code snippets that are related to one or multiple issues since we need to investigate the code to understand the potential design smell. 
Also, we have searched for publicly released datasets of faulty DL programs (including design issues and low performance) by checking replication packages of all published papers that studied problems in DL programs. Finally, we obtained four publicly available datasets of faulty DL programs gathered from SO and GitHub \cite{DL_bugs_1, DL_bugs_2, DL_faults, DL_fix2020}. All these studies investigated various faulty DL programs from SO and GitHub for their own research objectives including empirical study of bugs occurring in DL software systems written by TensorFlow, PyTorch and Caffe \cite{DL_bugs_1, DL_bugs_2}, proposing a taxonomy of real faults occurred in DL software systems \cite{DL_faults} and bug fix patterns in DL programs \cite{DL_fix2020}.  

For inspecting collected DL programs from either direct searching over SO or public datasets, we relied on certain inclusion and exclusion criteria 
to find relevant programs for identifying design smells:
\begin{itemize}
\item The program must have performance issues (e.g., low accuracy or detection precision),
\item The issue must not lead to program crash, hang or incorrect functionality. The program should be able to run and produce results,
\item The DL program must be developed using TensorFlow or Keras,
\item The DL model must be FNN,
\end{itemize}

This process left us with 659 DL programs to be analyzed. We have manually inspected all these artifacts to find relevant examples to identify design smells. We have used an open coding procedure \cite{seaman1999qualitative}. A shared document including the link to all artifacts have been used to make it possible for all authors to work together during the analysis. Each artifact was inspected by reading specific parts of its document (code snippet, comment, description) and all related discussion provided by the developer or other users (for samples from SO). Each sample was inspected by at least two of the authors to make sure that the root cause of the performance issue was a design inefficiency and was not related to generic programming faults or implementation issues. 

After analyzing all these data sources, we have derived a catalogue of 8 distinct design smells in deep FNN (a popular DL architecture). Since the arrangement of convolutions/poolings layers for extracting features and type/location of regularizers are two significant factors that affect the performance of deep FNNs, so we present the smells organised in two categories: Formation of the feature map and usage of regularization.

\subsection{Formation of the feature map, convolutions and poolings layers}
\begin{figure}
    \centering
    \includegraphics[width=0.9\linewidth]{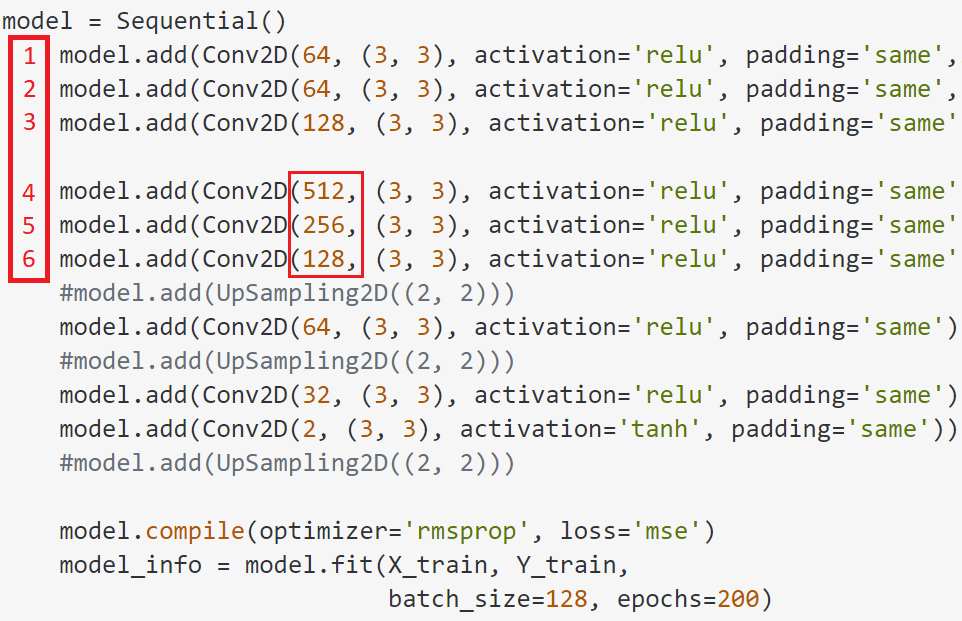}
    \caption{A part of DL program mentioned in SO\_50426349 as an example of design smell No. 1.}
    \label{fig:sample1}
    \vspace{-15pt}
\end{figure}

\begin{figure}
    \centering
    \includegraphics[width=0.75\linewidth]{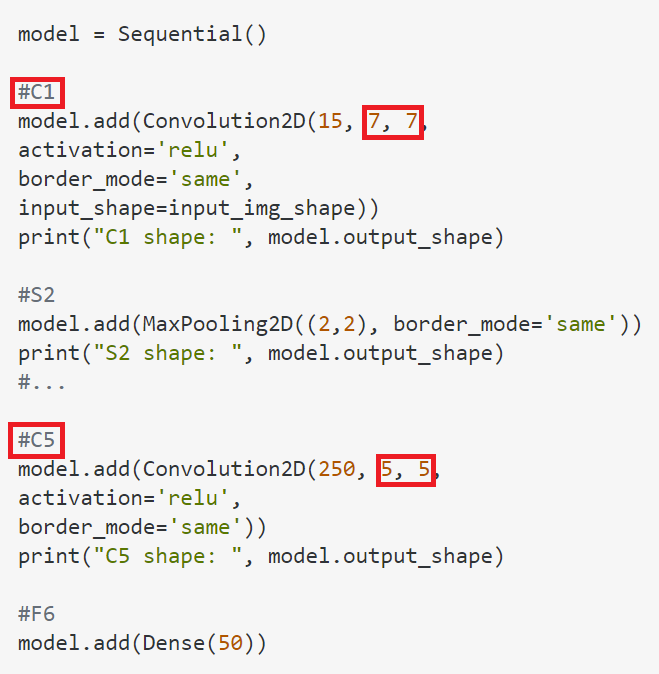}
    \caption{A part of DL model from SO\_38584268 as an example of design smell No. 2.}
    \label{fig:sample2}
\end{figure}

\begin{figure}
    \centering
    \includegraphics[width=.95\linewidth]{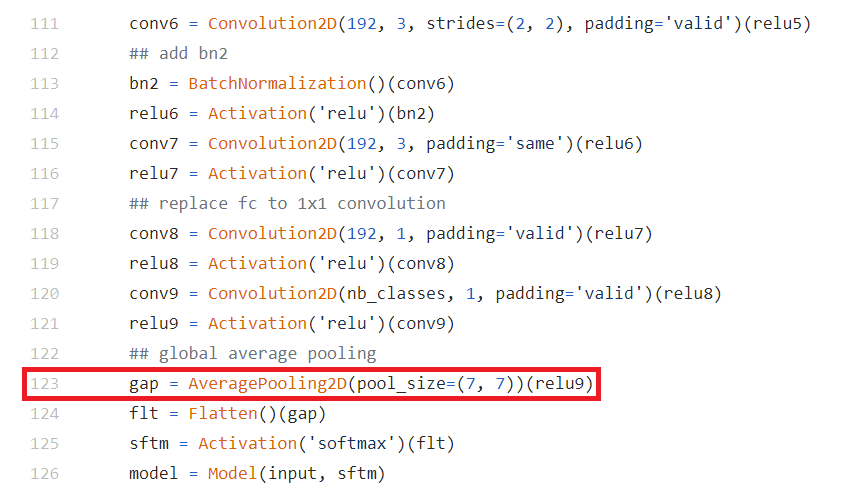}
    \caption{A part of DL program from GitHub as an example of design smell No. 5.}
    \label{fig:sample5}
\end{figure}

\textbf{Context:} Conventionally, a CNN architecture incorporates a bundle of convolutional layers with increasing filters count and separated by pooling layers to shrink gradually the feature map area. Hence, the extracted feature space tends to become deeper and narrower throughout the network until it becomes ready to be flatten and fed to the dense layers in charge of mapping the features into the target output.
\\ \\
\textbf{1. Non-expanding feature map}\\
\textbf{\textit{Bad smell description:}} A possible design mistake in CNNs is keeping the number of features the same (or even decrease it) as the architecture gets deeper. There should be a balance between retaining the detected features (and corresponding spatial relationship between them) and increasing the depth of the network \cite{depth_comp}.\\
\textbf{\textit{Consequences:}} If the developer fails to have a proper balance between the depth and size of the feature map, the overall performance would be negatively affected. While the stack of convolution and pooling layers extract and then compress the relevant feature map, if the architecture cannot increase the number of features, it will fail to deliver promising features to the dense layers.\\
\textbf{\textit{Recommended Refactoring:}} The number of feature maps should be gradually expanded while the feature map area is retracted. The growth of feature maps count is recommended \cite{depth_comp} to compensate the loss of representational expressiveness caused by the continuous decreasing of the spatial resolution of the learned feature maps. Therefore, throughout the layers, the feature space becomes synchronously narrower and deeper until it gets ready to be flatten and fed as input vector to the dense layers.\\
\textbf{\textit{Example:}} An example of this bad smell is illustrated in Fig. \ref{fig:sample1} extracted from SO post \#50426349. The developer did not grow the number of feature maps through layers 4 to 6. The number of layers and the size of 2-Dimensional convolution layers in the code snippet are highlighted in red.
\\ \\
\textbf{2. Losing local correlation}\\
\textbf{\textit{Bad smell description:}} In CNNs, promising features are extracted and then delivered to the dense layers by the stack of convolutional layers. For an effective feature extraction, setting proper window size for spatial filtering is crucial. If the developer does not grow the window size when the model gets deeper, the model will fail to extract the relevant features \cite{lecun2015deep}. Some developers start with a relatively large window size for spatial filtering and keep it the same for all convolutional layers which is a bad practice leading to loss of feature information. In fact, some developers only rely on the internal mechanism of convolutional and pooling layers for extracting relevant information without proper parameter settings/tuning.\\
\textbf{\textit{Consequences:}} If the model does not start with a relatively small window size (for gathering low-level information) and then grow the window size gradually (to extract high-level features), it will fail to extract useful features for the next processings. It makes sense that by using CNNs, the locality of information is crucial for performing the task. Thus, it is important to preserve locality throughout CNN to guarantee its success in detecting various features and relations between them \cite{lecun2015deep}. Furthermore, early convolutional layers learn lower level features while deeper ones learn more high-level and domain specific concepts.\\
\textbf{\textit{Recommended refactoring:}} The local window size for spatial filtering should generally increase or stay the same throughout the convolutional layers. It is recommended to start with small spatial filtering to collect much local information and then gradually increase it to represent more compound information \cite{VGGNet, szegedy2016rethinking}.\\
\textbf{\textit{Example:}} Fig. \ref{fig:sample2} shows a part of the code from SO post \#38584268 that defines a CNN with two convolutional layers 
The developer increased the kernel size (local window size) in successive convolution layers while should increase or at least keep it the same. The affected layers and corresponding API's arguments are marked in red in the code snippet.
\\ \\
\textbf{3. Heterogeneous blocks of CNNs}\\
\textbf{\textit{Bad smell description:}} Building a deeper model by only stacking a set convolution and pooling layers without appropriate configuration is a bad practice among DL developers. Even with proper adjustment of the number of features, the size of the local window, and the area of feature map along convolutional/pooling layers (as mentioned in the \textit{Non-expanding feature map} and the \textit{Losing local correlation} smells), efficient feature extraction can be affected by the lack of sufficient convolutional blocks \cite{he2016deep}. DL developers are used to define only one convolutional layer at each stage of a cascade of convolutional/pooling layers and increase the kernel size if it does not work properly. Depending on the application and the input data, usually, only one block of convolutional with large spatial filtering size at each stage is the 
minimum that the model needs to extract effective features efficiently.\\  
\textbf{\textit{Consequences:}} Only one convolutional block may not be enough for providing the required nonlinearity of feature extraction. On the other hand, large kernel sizes increase the computational burden significantly. As an example, recent NVIDIA cuDNN library (version 5.x or higher) is not optimized for larger kernels such as 5 × 5 and 7 × 7, whereas CNN with entirely 3 × 3 filters achieved a substantial boost in cuDNN performance \cite{cuDNNlink}.\\ 
\textbf{\textit{Recommended refactoring:}} Deep CNN should favor blocks of 2, 3, or even 4 homogeneous convolutional layers with similar characteristics. Advanced CNN architectures \cite{krizhevsky2012imagenet, he2016deep, iandola2014densenet} have shown the benefit of having several homogeneous groups of layers, where each one is specialized to achieve a particular goal. Indeed, building blocks of convolutional layers with similar characteristics (i.e., the same number of feature maps and feature map sizes) increases the homogeneity and the structure symmetry within the CNN. Hence, larger kernels can be replaced into a cascade of smaller ones, e.g., one 5 × 5 can be replaced by two 3 × 3, or four 2 × 2 kernels. Spatial filtering with reduced size enhances the nonlinearity and yields better accuracy \cite{VGGNet}. Moreover, it massively decreases the computation power requirement.\\
\\
\textbf{4. Too much down-sampling}\\
\textbf{\textit{Bad smell description:}} Usually DL developers define a pooling layer (down-sampling) after any convolutional layer. While down-sampling is inevitable in CNN models, it is not a good practice to perform the down-sampling right after each convolutional layer particularly for early layers.\\
\textbf{\textit{Consequences:}} Larger feature-maps, especially in the early layers, provide more valuable information for the CNN to utilize and improve its discriminative power \cite{he2015convolutional, szegedy2016rethinking, iandola2016squeezenet}. Therefore, it is crucial to avoid prematurely down-sampling and excessive appliance of pooling. Otherwise, the model will lose some information extracted in early layers resulting in poor performance.\\
\textbf{\textit{Recommended refactoring:}} Deep CNN should not apply pooling after every convolution. For instance, we use, as an approximation, the minimum of 10 layers to consider a CNN deep and 1/3 as threshold for the proportion of pooling layers with respect to the total of convolutional layers (convolution + pooling) to pinpoint a high amount of pooling.\\
\\
\textbf{5. Non-dominating down-sampling}\\
\textbf{\textit{Bad smell description:}} In fact, down-sampling \cite{strided_conv} in the cascade of CNNs can be done by max- or average-pooling or strided convolution (strides greater than 1). Using average-pooling is recognized as a bad design choice for CNN models \cite{MaxPooling_Sup}, particularly for image-like data.\\
\textbf{\textit{Consequences:}} Average-pooling ignores some invariances in data. Since extracting invariant features (those are not affected by scaling or various transformations) is crucial for image processing and object recognition, failure to deliver such features to the dense layers leads to an accuracy degradation of classification. Moreover, it can affect the generalization capability of the model.\\
\textbf{\textit{Recommended refactoring:}} Max-pooling is the preferred down-sampling strategy, so all the down-sampling is recommended to be changed to max-pooling. Max-pooling operation has been shown to be extremely superior for capturing invariances in data with spatial information, compared to other down-sampling operations \cite{MaxPooling_Sup}.\\
\textbf{\textit{Example:}} Fig. \ref{fig:sample5} illustrates a part of code from a GitHub repository\footnote{\url{https://github.com/yumatsuoka/comp_DNNfw/commit/30e0973892bc344aa17cd36a63dc61a062ad93e4}} as an example of this bad smell. It is highlighted in the code snippet that developer used average-pooling instead of recommended max-pooling.  

\subsection{Using regularization}
\begin{figure}
    \centering
    \includegraphics[width=0.75\linewidth]{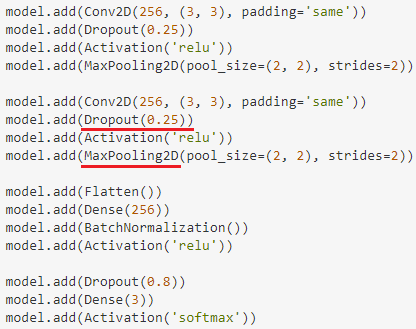}
    \caption{A part of DL program mentioned in SO\_60566498 as an example of design smell No. 6.}
    \label{fig:sample6}
\end{figure}

\begin{figure}
    \centering
    \includegraphics[width=0.75\linewidth]{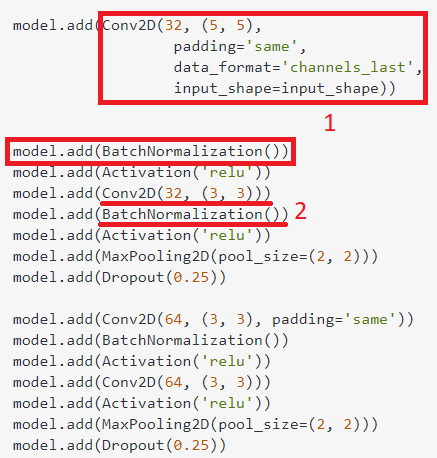}
    \caption{A part of DL program mentioned in SO\_49117607 as an example of design smell No. 7.}
    \label{fig:sample7}
\end{figure}

\textbf{Context:} Order and combination of regularization can affect the performance of FNN significantly \cite{batchnorm, disharmony_dropout_batchnorm, systematic_CNNs}. Moreover, the regularization functionality may interfere with other FNN’s components. Therefore, regularization should be used properly (place, order and combination) to ensure their effectiveness. The following smells discuss bad practices on the usage of regularizations in a FNN architecture.\\
\\
\textbf{6. Useless Dropout}\\
\textbf{\textit{Bad smell description:}} It is well-known among DL developers that dropout helps to avoid overfitting, however, using it before down-sampling layers will counteract its effect \cite{systematic_CNNs}.\\
\textbf{\textit{Consequences:}} Dropping out the activation before the pooling could have no effect except in cases where the masked units correspond to maximums within input pooling windows. The reason is that the max-pooling keeps only these maximums as inputs for next layers. With the neutralized dropouts, the model will suffer from overfitting and poor performance.\\
\textbf{\textit{Recommended refactoring:}} Dropout layer must be placed after the maximum pooling layer to be more effective. Considering the case studies with max-pooling layers \cite{dropout}, the dropout has been applied on the pooled feature maps, which becomes a heuristic followed by the state-of-the-art CNN architectures \cite{systematic_CNNs, practical_CNNs}.\\
\textbf{\textit{Example:}} In the example shown in Fig. \ref{fig:sample6}, extracted from SO post \#60566498, the developer has used “Dropout” before “MaxPooling2D” (both underlined by red in the code). The developer complained about increasing validation loss and bad performance of his model in the post.\\ 
\\
\textbf{7. Bias with Batchnorm}\\
\textbf{\textit{Bad smell description:}} Normally learning layers in FNN benefits from bias with different initializations. When using batchnorm, keeping bias values in layers is not a good practice \cite{batchnorm}.\\
\textbf{\textit{Consequences:}} Actually, the effect of batchnorm will be diminished in the presence of a bias. Batchnorm applies, after the normalization, a linear transformation to scale and shift the normalized activations $\hat{a} = \alpha a + \beta$, where $\alpha$ and $\beta$ are learnable parameters. This allows DNN to compensate for any loss of information by the value distortions in order to preserve its expressive power. Since, batchnorm already adds a $\beta$ term fulfilling the same role of bias, “its effect will be canceled” \cite{batchnorm} in the presence of a bias.\\
\textbf{\textit{Recommended refactoring:}} The bias should be removed or ignored in a learning layer that is equipped with a batchnorm.\\
\textbf{\textit{Example:}} The code snippet in Fig. \ref{fig:sample7}, extracted from SO post \#49117607, shows that the developer has used two learning layers (“Conv2D") without turning off the bias along with Batchnorm (both underlined by red in the code with 1 and 2 respectively).\\
\\
\textbf{8. Non-representative Statistics Estimation}\\
\textbf{\textit{Bad smell description:}} Another bad practice regarding regularizations is using batchnorm after dropout. The developers usually use different regularization techniques to maintain and improve performance of DL, however, they should be careful about the internal mechanism and effects of these two different regularization techniques \cite{disharmony_dropout_batchnorm}.\\
\textbf{\textit{Consequences:}} If the batchnorm is placed after the dropout, it will compute non-representative global statistics (i.e., moving average and moving variance) on the dropped outputs of the layer. Li et al. \cite{disharmony_dropout_batchnorm} discussed the effects of this disharmony between dropout and batchnorm and showed experimental results asserting their explanation.\\
\textbf{\textit{Recommended refactoring:}} Batchnorm should be applied before dropout. Therefore, a substitution in the model design is recommended if batchnorm is applied after dropout to address the issue.\\
\textbf{Example:} Fig. \ref{fig:sample7} illustrates a part of program presented in SO post \#55776436, showing that “Dropout" has been used before the “BatchNormalization" (a red box indicates affected lines and they are highlighted both with 1 and 2 respectively). The developer in his post complained about low classification accuracy.
\begin{figure}
    \centering
    \includegraphics[width=0.9\linewidth]{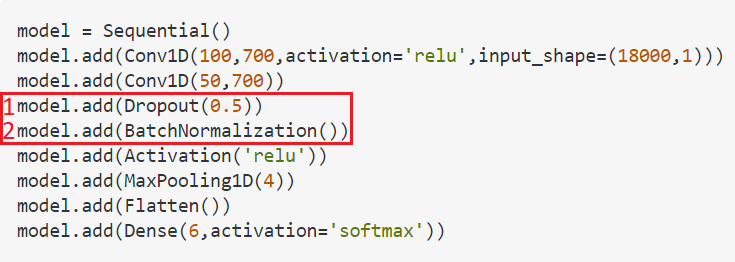}
    \caption{A part of DL program mentioned in SO\_55776436 as an example of design smell No. 8.}
    \label{fig:sample8}
\end{figure}

\section{Relevance Assessment of Design Smells}\label{survey}
After identifying bad design smells in DL models, we wanted to assess them. Our goal was to know whether developers/researchers evaluate them as relevant and possibly worthwhile to be addressed. Hence, we run a survey to validate our catalogue of DL design smells and collect views of DL developers/researchers about it. In the following, first the methodology followed to conduct the survey is explained, then the results are presented.
\subsection{Survey Design}
Our survey was created using Google Forms \cite{googleForm}, a well-known online tool for creating and sharing online surveys and quizzes. The survey is organized in three parts. In the first part, we ask some demographic questions about the participant: i) their role in the organization or job title (e.g., developer, researcher, student), ii) their number of years of work/research experience in ML/DL and iii) their used programming languages/frameworks. The second part asks specific questions about the design smells. We provide a description for each of our 8 design smells and a multiple-choice question asking the participant about the perceived relevance of the smell. The participant is instructed to provide a score on a 5-level Likert scale \cite{oppenheim2000questionnaire}. Moreover, for each question, we provide an open comment box to the participants, asking for their feedback about the definition of the design smell. 
In the final part, we ask (i) if the participant has observed any other frequent/significant design issues that have not been considered in our survey. (ii) We also ask them if a tool for detecting such smells would be useful or not, and (iii) whether they would opt for using such tool. We ask this last question because one could find a tool useful, but more for others (
like junior developers/researches) than for themselves. At the end of the survey, we provided an open comment box allowing participants to 
share any additional comments (that they wished) with us.

The target group of candidates for this survey is developers, practitioners, or researchers with a good experience in DL and particularly in FNNs. The first group of candidates was derived from authors’ personal contacts, actually 16 experts. The second group of candidates came from GitHub. To find participants with a good understanding of FNNs over GitHub, we used its REST APIs \cite{githubREST}. First, we identified the relevant repositories that include “feedforward neural networks” and “convolutional neural networks” in their description. We excluded repositories that were not active since 2019. Finally, we extracted active contributors’ emails from 12192 selected repositories. This process left us with 3650 unique email addresses and we successfully distributed the survey participation request to 3605 email addresses. The third group of candidates came from \textit{Reddit}. To recruit participants, the questionnaire was posted on two relevant Reddit channels: \textit{deeplearning} and \textit{MachineLearning}. When sending/posting the questionnaire, we explained the purpose, scope and the estimated participation duration (5-10 minutes) of the survey in a quick message. Moreover, we asserted that the survey is kept anonymous, but the respondents were able to provide their emails for further communication and receiving a summary of the study.
\subsection{Validation results}
\begin{figure*}
    \centering
   \includegraphics[width=.95\linewidth]{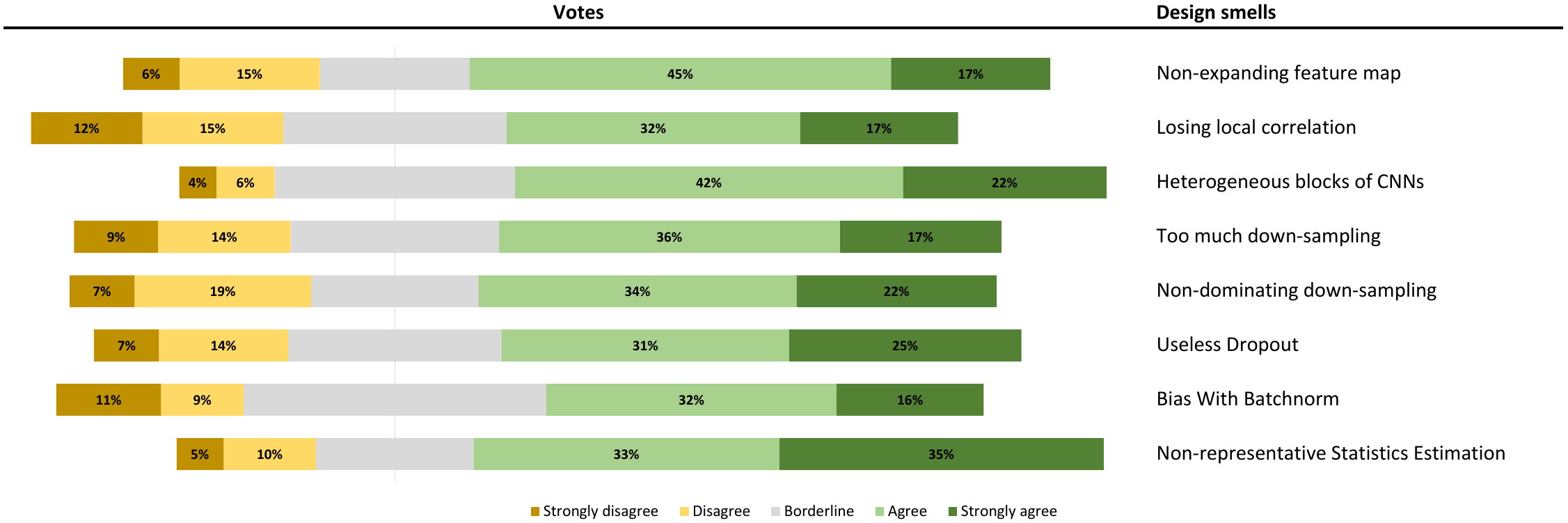}
    \caption{Validation results: Perceived relevance of the 8 design smells} 
    \label{fig:result1}
\end{figure*}

The survey was open for three weeks resulting in 81 responses in total. Regarding our question on work/research experience in DL, 20 respondents had less than 1 year experience, 41 between 1 and 3 years, 10 between 3 and 5 years, and 10 had more than 5 years. Almost all of the respondents (80 of 81) were using Python for DL development and only one indicated C++ as his favorite programming language. Among DL frameworks, TensorFlow was the most popular one with 59 votes. Keras and PyTorch received 45 and 42 votes respectively. 
Fig. \ref{fig:result1} shows the results of relevance assessment for 8 identified smells in the form of diverging stacked bar charts. Dark/light green color indicates the proportion of “Strongly agree” and “Agree” responses, while dark/light brown indicates the proportion of “Strongly disagree” and “Disagree” responses. \textit{Non-representative Statistics Estimation} is the most popular smell in our survey as it received 68\% of positive votes (“Strongly agree” and “Agree”) while \textit{Bias With Batchnorm} received the minimum positive rate of 47\%. On the other hand, the  highest negative feedback (“Strongly disagree” and “Disagree”) was recorded for \textit{Losing local correlation} with 27\%. In the following, we discuss the validation results and received comments for each smell.\\
\textbf{1. Non-expanding feature map:} In general, respondents agree (about 63\% of positive responses: “Strongly agree” and “Agree”) that keeping the number of features the same (or even decrease it) as the architecture gets deeper is a design mistake in DL models, e.g., one commented that: \textit{“I strongly agree with this statement. The number of channels must be increased so as to capture more complex features which appear as the layers grow deeper”}. However, there are some neutral and negative responses. Some of them asserted that this is the case only for classification tasks. Most of the negative/neutral comments explained that this design smell is not always true and the expansion of the feature map depends on data, application (task that DL model designed for) or network architecture. They used to consider the size of the feature map as a hyperparameter that should be tuned on the validation loss, e.g., \textit{“According to me the size of feature map is a hyperparameter and will depend on the size of the network (Depth) hence I neither agree or disagree with the given statement, since sometimes a combination of small and larger feature maps work well like in inception model.”}. Another respondent mentioned that s/he preferred to see an only slightly decreasing number of information processing units as the model gets deeper, and if the number of points is quartered (e.g., by max-pooling), the number of feature channels should be doubled or tripled.

\textbf{2. Losing local correlation:} This smell receives a low positive response rate of 49\%, the highest negative feedback among all smells (27\%: “Strongly disagree” and “Disagree”) from respondents and 24\% of neutral responses. While respondents agree that the window size is an important factor and should be adjusted as the network gets deeper (e.g., \textit{“I agree with this statement however increasing the window size will slow the training but our aim for a better model is achieved”}), they believe that non-growing window size across the network is not always a bad practice (e.g., \textit{“I think the windows size for spatial filtering should be directly proportional to how deep the network’s layers are”}). They mentioned that there are plenty of simple applications where fixing a window size is enough to achieve a reasonable performance and this approach makes implementation easier and hyperparameter tuning simpler (e.g., \textit{“The models I’ve worked with are all relatively small but I kept the window size the same, it worked fine”}). There are comments stating that if we start by a small dimension and grow it, we may have false correlation as a result of the larger subsequent layers in some cases. Another respondent rephrased our statement as \textit{“start with and keep (or slightly grow) a small window size”}. Three other comments mentioned autoencoder networks (since they benefit from CNNs) by stating that this characteristic is observed on the second half (decoder) of autoencoders but not in the first half, so this design smell can be true or false depending on context. From neutral responses, we have: \textit{“I have seen a case where first a large spatial filter after that constant filter size provided more performance than gradually increasing filter size in a larger CNN model. Though I have also seen the logic above working well”}.

\textbf{3. Heterogeneous blocks of CNNs:} Respondents have an agreement (64\%) with soundness and prevalence of this smell. Also, it received the minimum negative response of 10\% in our survey. They believed that we need multiple symmetric blocks of CNNs for effective feature extraction particularly in large models with enough depth not in small or medium ones. It was acknowledged that multiple layers are needed, not only to map complex relationships but also to be able to generate a sufficiently large receptive field: \textit{“a higher representation level is obtained with every additional convolutional layer”}. However, we received opposite views mentioning different aspects. Some experts commented that the designer should not spend too much effort on interpreting the activity of a single block and not try to set a goal for each block a priori, for example: \textit{“I agree with your claim except the last sentence”}. Others stated that convolutional blocks may be made of a single, several homogeneous or heterogeneous ones, and the design choice depends on the application: \textit{“the network size is determined primarily by the dataset size”}.

\textbf{4. Too much down-sampling:} More than half of respondents vote positively for this case (56\%), and the same proportion vote neutrally and negatively (22\%). We observed an agreement on the necessity of a balance between down-sampling vs. feature detection and not using too much down-sampling (\textit{“Too much down sampling can provide rigged results”} or \textit{“You do want to avoid downsampling too much, mostly because you're going to bottleneck all your information to nothing”}). However, controversial opinions are on accepting it as a rule and on the suggested 1/3 threshold. Some comments mentioned that there is no fix ratio and the optimum ratio that fits perfectly could be achieved by hyperparameter tuning, for example: \textit{“but I've seen optimal architectures in which that ratio is much higher (e.g.: 1:1) as well as much lower (e.g.: 1:10)”} or \textit{“I think it would be difficult to prove such rules apply to every CNN and every problem domain. Also, I have seen and used CNNs with no down-sampling layers”}. Another respondent mentioned that hesitancy to down-sample may increase CNN processing time while mostly preserving “junk” data in the network so the designer should be careful about it.

\textbf{5. Non-dominating down-sampling:} Similar to the previous smell, there is a marginal agreement on this one by 56\% of positive responses. Moreover, this case received a substantial rate of negative reactions, i.e., 26\%. According to the submitted comments, respondents acknowledged max-pooling as a dominant choice in most cases supported by results-driven (e.g., natural image data) and neuroscience-driven arguments. However, this is not the case always: \textit{“max pooling proves better than avg pooling but it cannot be completely ruled out”}, \textit{“Indiscriminate use of average pooling may suggest a code smell”} or \textit{“the decision I would say should be based on what features are being extracted and what is the model trying to learn”}. They mentioned that for some applications like extraction of a global parameter from an image, average-pooling can be more useful. Another respondent suggested using average-pooling instead of max-pooling in Generative Adversarial Networks (GAN) to avoid sparse loss. Finally, we found this comment very helpful: \textit{“Although contrast is a good way to see things, nuance is important. Nuance is lost with max-pooling especially with aggressive down-sampling or at later layers”}.

\textbf{6. Useless Dropout:} According to received responses, 56\% of respondents indicate their agreement with this smell. Although there were some strong positive comments like: \textit{“I generally don't include dropout before pooling”} or \textit{“ it's a rough heuristic to keep dropouts after pooling but it works well”}, negative responses expressed two main points against the statement of the smell: 1) type of dropout: element-wise vs. feature-wise, and 2) its effectinevess compared to batchnorm. Three respondents proposed that feature-wise dropout (dropping some proportion of feature maps rather than pixels or spatial dropout) should be more effective than random dropout for most applications by considering that \textit{“it does not matter at all whether it's used before or after pooling (since entire feature maps are dropped)”}. Two others suggested that dropout was being deprecated by batchnorm.

\textbf{7. Bias With Batchnorm:} Less than half of respondents went positively with this smell (47\%) while it received the most neutral votes in our survey by 33\%. Responders with positive votes stated that using bias with batchnorm is a bad practice and they avoid it generally. By reviewing comments, we come to the conclusion that negative and neutral voters believed that using bias with batchnorm is not harmful: \textit{“The conv bias is redundant with the BN bias, but I don't think it's harmful to keep it (just wasteful)”}, \textit{“I cannot see the presence of bias nodes being a problem”} or \textit{“the additional bias will simply "cancel" and the same representation is learned anyway”}. Therefore, the design smell does not look wrong and avoiding it can be helpful at least for keeping the model simpler.

\textbf{8. Non-representative Statistics Estimation:} There is a general agreement in this case since we received 68\% of positive votes as the most popular smell in our survey. A majority of respondents believed that using batchnorm after dropout would lead to non-representative statistics: \textit{“if batch normalisation is done after dropout then it will normalise the output coming after dropping the some connection (nodes)”}. However, there were also some negative comments on the smell. The main criticism was that the order of batchnorm and dropout does not have a significant impact on the performance of a DL model.

\begin{figure}
    \centering
   \includegraphics[width=0.9\linewidth]{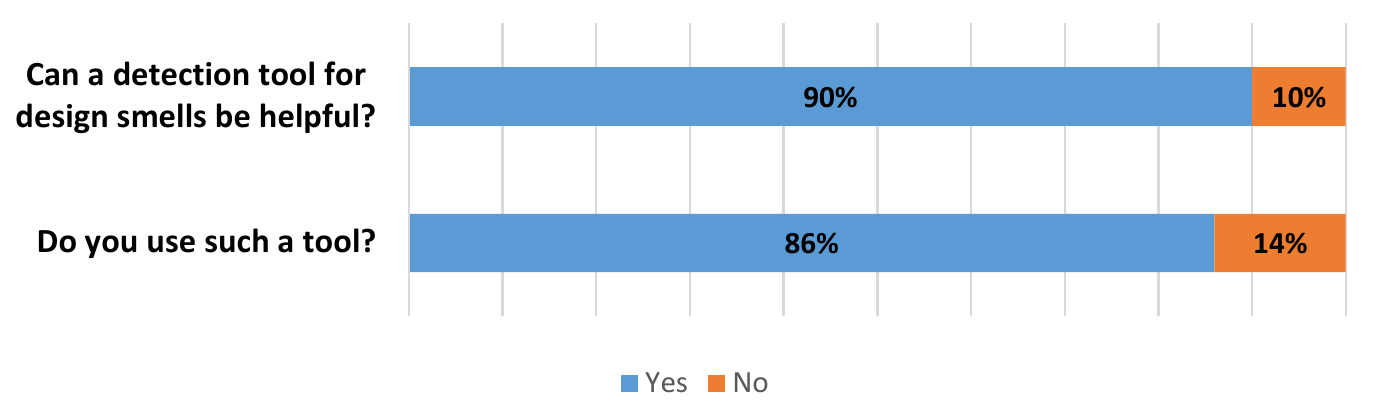}
    \caption{Survey results about a detection tool}
    \label{fig:result2}
\end{figure}

The results of our questions about the usefulness of a potential tool for detecting the identified smells are shown in Fig. \ref{fig:result2}. 
A significant majority of respondents, actually 90\%, expressed a positive opinion for such a detection tool. Our follow-up question regarding whether they would use this tool if it became available, received another high positive reaction rate of 86\%. We attribute the slight drop to some experienced respondents recognizing that a detection tool would be useful but not necessary to them. 
Finally, all respondents surprisingly answered our question about other frequent/significant smells not considered in this survey and further identification of smells. They suggested the investigation of potential design smells related to various components of DL programs, including: (i) Initialization methods, (ii) Other architectures like fully and autoencoder CNNs, (iii) Some hyperparameter: like learning rate for different layers, (iv) The choice and location of activation functions, (v) Attention layers, (vi) Transfer learning.

\subsection{Discussion}
Among the comments received in our survey, some respondents mentioned that although the proposed design smells have stated promising points for sketching DL models, hyperparameter tuning is inevitable after any initial design and the model’s performance can be improved significantly by a proper hyperparameter search, for example: \textit{“... just set up your hypermodel to accept these as tunable parameters and search the space”} or \textit{“... allowing users to perform a flexible hyperparameter to fit the model to their particular needs”}. They stated that given the range of applications for DL, many design/configuration choices are domain-, data- and preprocessing-dependent. Therefore, experiments (including for hyperparameter tuning) may be required in some cases to identify the issues. However, we believe that having a catalogue of known bad practices while designing DL models, will help developers to avoid smells in their models. Even if the proposed smells do not cover all domains, they are still useful for the covered architecture/domains. Moreover, avoiding those smells will save time, effort and computational resources during test or hyperparameter tuning. 

\section{Threats to Validity}\label{threats}
First of all, threats to construct validity may affect the relevance of the identified design smells which is assessed by a survey. In our survey, respondents were requested to indicate the perceived significance of smells described by a short explanation of the problem/situation. We have used relevant terminology and provided technical details in our descriptions to address this threat. Moreover, respondents were able to mention comments for each smell in the survey and we have not observed any comment complaining about possible misunderstanding in the description or context. It is also possible that our descriptions in the survey affected participant’s view directing them toward our proposed design smells. To address this concern, we asked participants at the end of our survey to freely comment on missing issues in our study.

There are internal threats to the validity of this research that may affect its achievements. The identification of design smells could be biased during reviewing previous works and manual inspection of artifacts. To address this issue, a clear systematic approach is followed in our study. We have investigated only ``closed" issues from GitHub and questions with ``at least one accepted" answer from SO; ensuring that we analyzed only issues that were solved. Moreover, participants in the survey have not been involved in the process of identifying smells and have different levels of expertise/background. Although the catalogue was prepared using DL programs developed by two popular frameworks of TensorFlow and Keras, we kept the title and description of the smells as general as possible and we believe that they are helpful for developers/researchers working with other frameworks as well.
 
External validity threats may impact the generalization of our findings. We indeed are aware that the proposed catalogue is not complete. Since our paper is a first step in identifying design smells in DL programs, further studies are required to comprehensively investigate design smells in DL programs utilizing various structures. Furthermore, some smells can be extended in future work since currently they are specified for particular cases.

\section{Conclusion}\label{conclusion}
In this paper, we have specified 8 design smells in DL programs. Due to the prevalence and effectiveness of deep CNNs in real-world applications (particularly with image-like data), we have focused on this architecture. Basically, these smells are structural inefficiencies in DL models, that affect the performance of DL programs. We evaluated the validity and relevance of this catalogue by running a survey with 81 DL developers/researchers. 
In general, the developers perceived the proposed design smells as reflective of design or implementation problems, with  agreement  levels varying between 
47\% and 68\%. The analysis of the multiple comments received for each of the smells, indicates that almost all the design smells are found to be relevant and helpful by respondents. Many of the survey respondents encountered similar design issues described by the smells.

There are several directions for future work. First, we plan to introduce a detection tool for the proposed smells. An automatic method for finding design smells in DL programs will help developers to improve their DL models prior to deployment. Second, we plan to generalize some of the already identified smells to cover other contexts. 
Finally, a more comprehensive variety of smells can be proposed by covering other DL architectures. 

\balance
\bibliography{references}
\bibliographystyle{ieeetr}


\end{document}